\documentclass[icol, referee, pdflatex, sn-aps]{sn-jnl}

\usepackage{graphicx}%
\usepackage{multirow}%
\usepackage{amsmath,amssymb,amsfonts}%
\usepackage{amsthm}%
\usepackage{mathrsfs}%
\usepackage[title]{appendix}%
\usepackage{xcolor}%
\usepackage{textcomp}%
\usepackage{manyfoot}%
\usepackage{booktabs}%
\usepackage{algorithm}%
\usepackage{algorithmicx}%
\usepackage{algpseudocode}%
\usepackage{listings}%

\theoremstyle{thmstyleone}%
\theoremstyle{thmstyletwo}%

\theoremstyle{thmstylethree}%

\begin{document}

\title[Article Title]{Unraveling the magnetic and electronic complexity of intermetallic ErPd$_2$Si$_2$: Anisotropic thermal expansion, phase transitions, and twofold magnetotransport behavior}




\author[1]{\fnm{} \sur{Kaitong Sun}}
\author*[2]{\sur{Si Wu}}\email{wusi@nbu.edu.cn}
\author[1]{\fnm{} \sur{Guanping Xu}}
\author[3]{\fnm{} \sur{Lingwei Li}}
\author[4]{\fnm{} \sur{Hongyu Chen}}
\author[1]{\fnm{} \sur{Qian Zhao}}
\author[1]{\fnm{} \sur{Muqing Su}}
\author[5]{\fnm{} \sur{Wolfgang Schmidt}}
\author*[6]{\sur{Chongde Cao}}\email{caocd@nwpu.edu.cn}
\author*[1]{\sur{Hai-Feng Li}}\email{haifengli@um.edu.mo}
\affil[1]{\orgdiv{Institute of Applied Physics and Materials Engineering}, \orgname{University of Macau}, \orgaddress{\street{Avenida da Universidade}, \city{Taipa}, \postcode{999078}, \state{Macao SAR}, \country{China}}}
\affil[2]{\orgdiv{School of Physical Science and Technology}, \orgname{Ningbo University}, \orgaddress{\city{Ningbo}, \postcode{315211}, \country{China}}}
\affil[3]{\orgdiv{Key Laboratory of Novel Materials for Sensor of Zhejiang Province}, \orgname{Hangzhou Dianzi University}, \orgaddress{\city{Hangzhou}, \postcode{310018}, \country{China}}}
\affil[4]{\orgdiv{College of Mechanical Engineering}, \orgname{Zhejiang University of Technology}, \orgaddress{\city{Hangzhou}, \postcode{310023}, \country{China}}}
\affil[5]{\orgdiv{Forschungszentrum J$\ddot{\textrm{u}}$lich GmbH}, \orgname{J$\ddot{\textrm{u}}$lich Centre for Neutron Science at ILL, 71 avenue des Martyrs}, \orgaddress{\city{Grenoble}, \postcode{38042}, \country{France}}}
\affil[6]{\orgdiv{School of Physical Science and Technology}, \orgname{Northwestern Polytechnical University}, \orgaddress{\city{Xian}, \postcode{710072}, \country{China}}}


\abstract{We present a comprehensive investigation into the physical properties of intermetallic ErPd$_2$Si$_2$, a compound renowned for its intriguing magnetic and electronic characteristics. We confirm the tetragonal crystal structure of ErPd$_2$Si$_2$ within the $I4/mmm$ space group. Notably, we observed anisotropic thermal expansion, with the lattice constant $a$ expanding and $c$ contracting between 15 K and 300 K. This behavior is attributed to lattice vibrations and electronic contributions. Heat capacity measurements revealed three distinct temperature regimes: $T_1 \sim 3.0$ K, $T_\textrm{N} \sim 4.20$ K, and $T_2 \sim 15.31$ K. These correspond to the disappearance of spin-density waves, the onset of an incommensurate antiferromagnetic (AFM) structure, and the crystal-field splitting and/or the presence of short-range spin fluctuations, respectively. Remarkably, the AFM phase transition anomaly was observed exclusively in low-field magnetization data (120 Oe) at $T_\textrm{N}$. A high magnetic field ($B =$ 3 T) effectively suppressed this anomaly, likely due to spin-flop and spin-flip transitions. Furthermore, the extracted effective PM moments closely matched the expected theoretical value, suggesting a dominant magnetic contribution from localized 4$f$ spins of Er. Additionally, significant differences in resistance ($R$) values at low temperatures under applied $B$ indicated a magnetoresistance (MR) effect with a minimum value of -4.36\%. Notably, the measured MR effect exhibited anisotropic behavior, where changes in the strength or direction of the applied $B$ induced variations in the MR effect. A twofold symmetry of $R$ was discerned at 3 T and 9 T, originating from the orientation of spin moments relative to the applied $B$. Intriguingly, above $T_\textrm{N}$, short-range spin fluctuations also displayed a preferred orientation along the $c$-axis due to single-ion anisotropy. Moreover, the $R$ demonstrated a clear $B$ dependence below 30 K. The magnetic-field point where $R$ transitions from linear $B$ dependence to a stable state increased with temperature: $\sim$ 3 T (at 2 K), $\sim$ 4.5 T (at 4 K), and $\sim$ 6 T (at 10 K). Our study sheds light on the magnetic and electronic properties of ErPd$_2$Si$_2$, offering valuable insights for potential applications in spintronics and quantum technologies.}

\keywords{Anisotropic thermal expansion, magnetoresistance, antiferromagnetic phase transitions, electronic transport properties, and magnetotransport behavior}



\maketitle

\section{Introduction}

High-temperature superconductivity consistently emerges in quasi-layered structures upon introducing electrons and holes through doping. A comprehensive understanding of the pairing mechanism in high-$T_c$ superconductors necessitates a thorough grasp of the interplay between localized antiferromagnetic (AFM) spins and itinerant electrons \cite{MAGGIORA20241}. Angular-dependent magnetoresistance (AMR), observed through resistivity measurements with an external magnetic field rotating along the single-crystal easy magnetization axis, provides evidence of spin and electron interactions \cite{ZhenZhao77406}. For instance, in the parent compound of the iron-based superconductor SrFe$_2$As$_2$, AMR exhibits twofold symmetry oscillation within the $ab$ plane, consistent with a strip-type spin structure below the spin-density-wave (SDW) temperature \cite{chen2008transport}. Similarly, in the cuprate superconductor Nd$_{2-x}$Ce$_x$CuO$_4$, AMR displays fourfold symmetry in noncollinear AFM phases and twofold symmetry at the spin reorientation transition temperature \cite{chen2005thermal}. Additionally, in the cobalt-oxide superconductor Na$_{0.48}$CoO$_2$, charge strip formation induces twofold AMR oscillation \cite{hu2006angular}. This demonstrates the importance of AMR in understanding the spin and electron interaction in various high-temperature superconductors.

ErPd$_2$Si$_2$, akin to iron-based superconductors, possesses a smaller interlayer spacing \cite{yan2008structural, cao2014single} and has garnered attention for its magnetic and electronic properties. As a member of the rare-earth intermetallic compounds with a ThCr$_2$Si$_2$-type crystal structure, ErPd$_2$Si$_2$ exhibits a complex interplay between magnetic ordering, electronic transport, and lattice dynamics \cite{MAZILU2008221, cao2014single}. The inclusion of the rare-earth element Erbium (Er) introduces localized 4$f$ electron states \cite{PhysRevB.88.174517}, leading to intricate magnetic phenomena. ErPd$_2$Si$_2$ undergoes intriguing magnetic phase transitions, including the emergence of SDW, the establishment of a complex modulated AFM structure with Er moments aligned along the $c$-axis below $\sim$ 4.8 K \cite{tomala1994magnetic, Li2015-2}, and potential short-range spin fluctuations \cite{sampathkumaran2008magnetic}. These transitions could manifest as anomalies in properties such as heat capacity, magnetization, and electrical resistivity, underscoring the importance of investigating their nature and impact.

While ErPd$_2$Si$_2$ typically exhibits metallic behavior in its electrical properties, the origins of its magnetotransport phenomena, including magnetoresistance effects, anisotropic electron scattering under applied magnetic fields, and thermal expansion with temperature, remain unclear. Understanding these phenomena is crucial to uncovering the complex magnetic and electronic properties of ErPd$_2$Si$_2$, thereby expanding its potential for future technological applications. Therefore, this study aims to address these gaps through a detailed investigation of its physical properties.

In this paper, we present a comprehensive investigation into the physical properties of ErPd$_2$Si$_2$. These include structural characterization, heat capacity measurements, magnetization studies, and a detailed analysis of magnetoresistance behaviors. Notably, the angular dependence of in-plane and out-of-plane magnetoresistances exhibits twofold symmetry, providing transport evidence for the formation of a stripe-type antiferromagnetic (AFM) structure of Er moments in the spin-density wave (SDW) state. Our study builds on previous investigations by offering detailed experimental evidence of anisotropic thermal expansion, phase transitions, and magnetotransport behavior. This work provides new insights into the interplay between localized and itinerant magnetism, which were not fully understood in earlier studies. The detailed analysis of angular-dependent magnetoresistance and its connection to spin configurations distinguishes this study from prior reports, advancing the understanding of spin-electron interactions in intermetallic compounds.

\section{Experimental}

\subsection{Single crystal growth and characterizations}

Previous phase diagram studies show that the ErPd$_2$Si$_2$ compound melts congruently at $\sim$ 1420 $^\circ$C \cite{MAZILU2008221}. The single crystals of the ErPd$_2$Si$_2$ compound were grown using a well-equipped laser-diode floating-zone furnace (Model: LD-FZ-5-200W-VPO-PC-UM) at the University of Macau, Macao, China \cite{wu2020super}. The single crystal growth procedure aligns with those previously documented in the literature \cite{cao2014single, sun2022temperature}.

We utilized the neutron Laue diffractometer, OrientExpress (located at ILL, France) \cite{Ouladdiaf2006}, to confirm that the studied crystal is single-crystalline. The OrientExpress diffractometer employs a CCD detector coupled with a neutron scintillator to monitor Bragg diffraction spots in the reciprocal space of the ErPd$_2$Si$_2$ compound.

A single crystal of the ErPd$_2$Si$_2$ compound was aligned using an in-house X-ray Laue diffractometer and cut into suitable pieces with determined crystallographic orientations along the $c$ axis or the $\langle$1 1 0$\rangle$ direction.

\subsection{Structural characterizations}

We carefully pulverized an ErPd$_2$Si$_2$ single crystal into powder for the structure study, employing an in-house X-ray powder diffractometer (Rigaku, SmartLab 9 kW). The powder diffractometer utilized Cu $K_{\alpha1}$ = 1.54056 {\AA} and $K_{\alpha2}$ = 1.54439 {\AA} with an intensity ($II$) ratio of $II_{K\alpha2}${/}$II_{K\alpha1} =$ 0.5 as the radiation. Diffraction patterns were collected at a voltage of 45 kV and a current of 200 mA. The measured 2$\theta$ range extended from 30 to 78$^{\circ}$ with a constant step size of 0.005$^{\circ}$. X-ray powder diffraction (XRPD) experiments were carried out in the temperature range from 15 to 300 K with an interval of 5 K to extract the temperature-dependent structural information of ErPd$_2$Si$_2$. The collected XRPD data were refined using the FULLPROF SUITE software \cite{rodriguez1993recent}. Initial crystallographic parameters were referenced to those reported in a room-temperature XRPD study of ErPd$_2$Si$_2$ \cite{Li2015-2}. We selected a pseudo-voigt function to simulate the Bragg peak shape and used linear interpolation between automatically-selected data points to calculate the background contribution. The refinement process involved adjusting parameters such as scale factor, zero shift, lattice constants, peak shape parameters, asymmetry, atomic positions, and isotropic thermal parameter ($B_{iso}$).

\subsection{Characterization of physical properties}

The heat capacity, magnetization, and electrical transport property were measured with a quantum design physical property measurement system (PPMS DynaCool instrument). Heat capacity measurements for a small piece of the ErPd$_2$Si$_2$ single crystal were conducted using a thermal relaxation calorimeter on the PPMS DynaCool from 1.8 to 35 K at 0 T. D.C. magnetization measurements were performed on a superconducting quantum interference device (SQUID) magnetometer in vibrating-sample magnetometry (VSM) mode under applied magnetic fields of 120 Oe and 3 T, ranging from 2 to 300 K. Magnetization was measured in two modes: one after cooling with zero magnetic field (ZFC), and the other under the applied magnetic field (FC).

The temperature dependence of the electrical resistance ($R$) of the aligned and cut ErPd$_2$Si$_2$ single crystal was measured from 2 to 300 K at 0 and 6 T when the current ($I$) and applied magnetic field ($B$) were parallel to either the $\langle$1 1 0$\rangle$ orientation or the $c$-axis. The electrical $R$ of the aligned and cut ErPd$_2$Si$_2$ single crystal as a function of $B$-rotating angle ($\gamma$: clockwise away from the $c$ axis, 0--360$^\circ$; $\theta$: clockwise away from the $\langle$1 1 0$\rangle$ direction, 0--360$^\circ$) was measured with the rotational electrical transport option (ETO) on our PPMS DynaCool at applied magnetic fields of 0, 3, and 9 T and temperatures of 2, 10, 30, 150, and 300 K. The applied-magnetic-field dependent $R$ of the aligned and cut ErPd$_2$Si$_2$ single crystal was measured at temperatures of 2, 4, 10, 30, 150, and 300 K when the current $I$ and applied magnetic field $B$ were along either the $\langle$1 1 0$\rangle$ orientation or the $c$-axis.

\section{Results and discussion}

\subsection{Neutron Laue study}

\begin{figure} [!t]
\centering \includegraphics[width=0.48\textwidth]{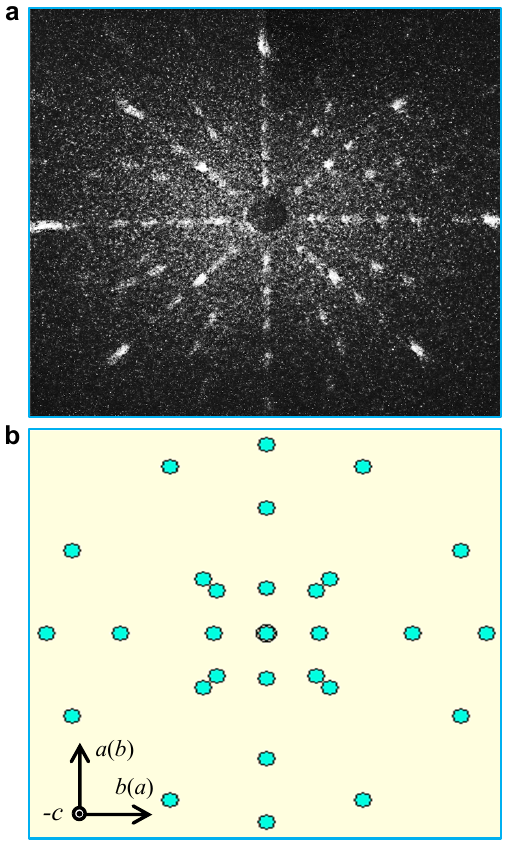}
\caption{
\textbf{a} A representative neutron Laue pattern of an ErPd$_2$Si$_2$ single crystal, with a detector-to-sample distance of 3.5 cm. A pinhole with a diameter of 3 mm was used for the neutron beam incident on the crystal, and the exposure time was set to 2 minutes. The incoming neutron beam is oriented perpendicular to the plane of the Laue spots and is thus parallel to the unit-cell edge, \emph{i.e.,} the neutron beam propagates along the crystallographic \emph{c}-axis. \textbf{b} The corresponding simulation of the neutron Laue pattern, with real-space lattice vectors indicated as labeled.}
\label{Laue}
\end{figure}

Figure~\ref{Laue}a illustrates the recorded back-diffraction Laue pattern, which maps the complete Miller indices in the (\emph{H} \emph{H} 0) scattering plane. The pattern exhibits four-fold symmetry, with sets of Laue spots arranged in lines or rings around the center. In this configuration, the \emph{c}-axis is perpendicular to the plane of the paper. Within the $I4/mmm$ space group of ErPd$_2$Si$_2$, the crystallographic \emph{a}-axis is equivalent to the \emph{b}-axis.

The recorded neutron Laue pattern was simulated using the software $\texttt{OrientExpress}$ \cite{Ouladdiaf2006}, as shown in Fig.~\ref{Laue}b, further validating the determination of the crystallographic orientations. This neutron Laue study confirms the single-crystalline nature of the grown ErPd$_2$Si$_2$ crystals.

\subsection{Structure study}

\begin{figure} [t]
\centering
\includegraphics[width = 0.82\textwidth] {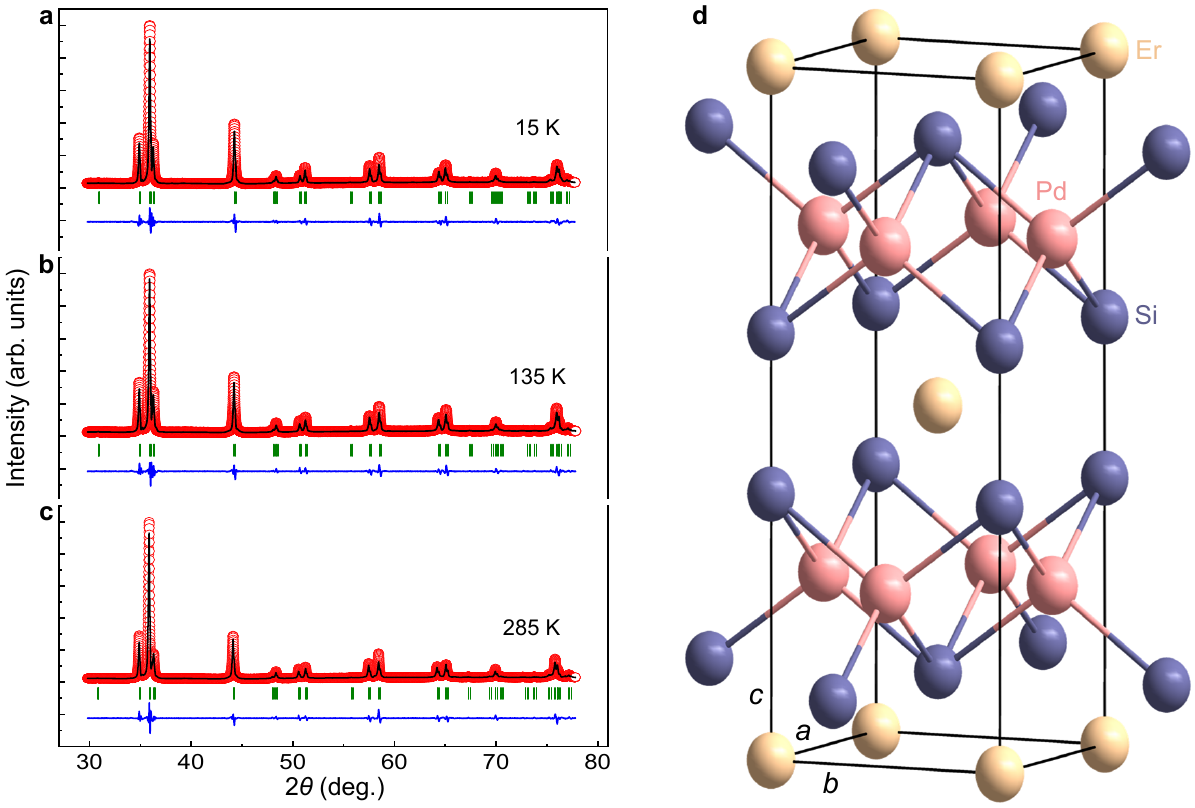}
\caption{Temperature-dependent X-ray powder diffraction patterns of a pulverized ErPd$_2$Si$_2$ single crystal, acquired at \textbf{a} 15 K, \textbf{b} 135 K, and \textbf{c} 285 K. The displayed data include the observed diffraction patterns (red circles) alongside the corresponding results of Rietveld refinement (black solid lines). The vertical bars (green) represent the positions of Bragg peaks. The lower solid lines (blue) show the difference between the observed and calculated patterns. \textbf{d} Crystal structure within a single unit cell of the ErPd$_2$Si$_2$ compound, characterized by a tetragonal system and the $I4/mmm$ space group.}
\label{TXRD}
\end{figure}

To investigate temperature-dependent structural characteristics, we conducted an XRPD study. Figure~\ref{TXRD}a-c illustrates three representative XRPD patterns obtained at temperatures of 15 K (Fig.~\ref{TXRD}a), 135 K (Fig.~\ref{TXRD}b), and 285 K (Fig.~\ref{TXRD}c). Accompanying Rietveld refinements are also presented. The refinement outcomes reveal that the ErPd$_2$Si$_2$ compound adopts a tetragonal crystal structure, with the observed Bragg peaks fitting well within the indexing parameters of the \emph{$I4/mmm$} space group (No. 139). Figure~\ref{TXRD}d depicts the resulting crystalline unit cell, highlighting the uniaxial coordination of Si along the $c$-axis, while the atomic positions of Er and Pd ions remain constant (Table~\ref{cellp}). Throughout the entire studied temperature range, we did not observe any Bragg peak splitting or additional peaks, particularly at high diffraction angles, implying the absence of structural phase transitions. The extracted crystallographic information at 15 K, 135 K, and 285 K is listed in Table~\ref{cellp}, where the low values of the reliability factors validate our refinements. These findings set the stage for a detailed examination of thermal expansions along the crystallographic $a$ and $c$ orientations.

\begin{figure} [t]
\centering
\includegraphics[width = 0.82\textwidth] {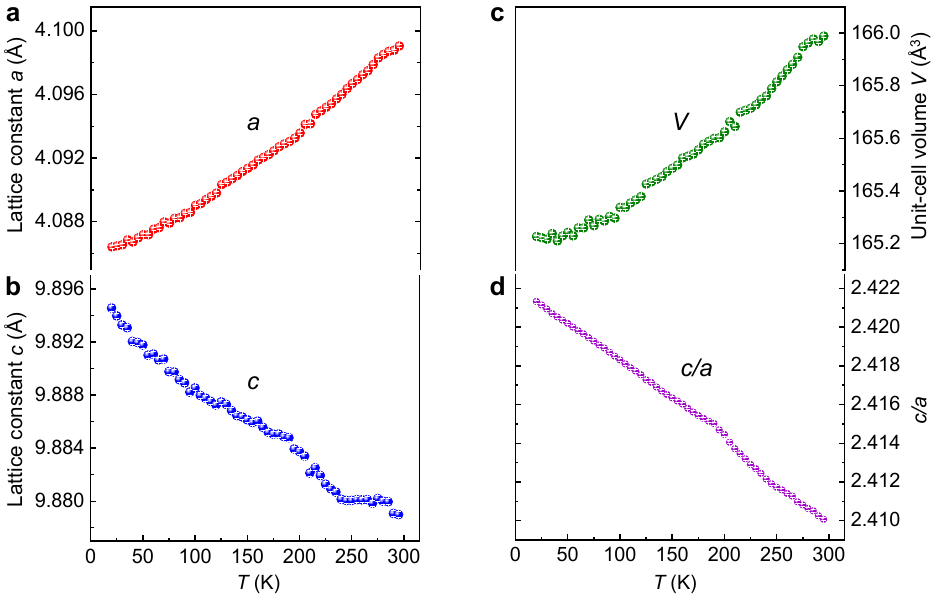}
\caption{Temperature-dependent structural parameters of ErPd$_2$Si$_2$ refined through XRPD analysis.
\textbf{a} Variation of lattice constant $a$ with temperature.
\textbf{b} Variation of lattice constant $c$ with temperature.
\textbf{c} Temperature-dependent changes in unit-cell volume $V$. The estimated standard deviations are derived from structure refinements in panels (\textbf{a}-\textbf{c}). \textbf{d} The ratio of lattice constants $c/a$ as a function of temperature.}
\label{latticep}
\end{figure}

To study thermal expansions along the crystallographic $a$ and $c$ orientations, a comprehensive temperature-dependent XRPD study was conducted. Figure~\ref{latticep} illustrates the temperature dependence of the lattice parameters for the ErPd$_2$Si$_2$ compound. As the temperature increases, the lattice constant $a$ expands (Fig.~\ref{latticep}a), while the lattice constant $c$ contracts (Fig.~\ref{latticep}b). The opposing temperature-dependent changes in lattice constants $a$ and $c$ collectively contribute to a positive thermal expansion in the unit-cell volume ($V$) (Fig.~\ref{latticep}c), with $(V_{\texttt{295K}} - V_{\texttt{20K}})/V_{\texttt{20K}}$ = 0.461(5)\%, approximately following the Gr$\ddot{\texttt{u}}$neisen rule at zero pressure in a first-order fashion \cite{C4RA06966H}. The anisotropic thermal expansion along the crystallographic directions is evident in the temperature-induced variations of lattice constants $a$ (positive) and $c$ (negative) between 20 K and 295 K, expressed as $(a_{\texttt{295K}} - a_{\texttt{20K}})/a_{\texttt{20K}}$ = 0.309(3)\% and $(c_{\texttt{295K}} - c_{\texttt{20K}})/c_{\texttt{20K}}$ = -0.157(3)\%. This comparison highlights the anisotropy in thermal expansion. Further analysis of Fig.~\ref{latticep}d reveals the temperature-dependent variation in the $c/a$ lattice constant ratio, where the error bars were calculated based on the propagation law of errors \cite{PhysRevB.102.019901}, providing additional evidence for the contraction of lattice constant $c$.

The ErPd$_2$Si$_2$ is an intermetallic compound. Its nonmagnetic (nonmag) thermal expansion arises from two main components, as described by the following formula:
\begin{eqnarray}
\varepsilon_{nonmag} = \varepsilon_{ele} + \varepsilon_{phon},
\label{Gr1}
\end{eqnarray}
where $\varepsilon_{ele} = KT^2$ represents the electronic contribution to the thermal expansion of the lattice parameter ($\varepsilon$). The phonon contribution ($\varepsilon_{phon}$) is calculated using the second-order Grüneisen rule at zero pressure \cite{li2012possible, Wallace1998, Vocadlo2002}:
\setlength\arraycolsep{1.4pt} 
\begin{eqnarray}
\varepsilon(T) = \varepsilon_0 + \varepsilon_0\frac{U}{Q-BU},
\label{Gr1}
\end{eqnarray}
where $\varepsilon_0$ is the lattice parameter at zero Kelvin, and the internal energy (\emph{U}) is estimated using the Debye approximation:
\begin{eqnarray}
U(T) = 9Nk_BT\left(\frac{T}{\Theta_D}\right)^3 \int^{\frac{\Theta_D}{T}}_0 \frac{x^3}{e^x - 1}dx,
\label{Gr2}
\end{eqnarray}
where \emph{N} is the number of atoms per formula unit, and $\Theta_D$ is the Debye temperature. In most cases, the electronic contribution ($\varepsilon_{ele}$) in Equ.~(\ref{Gr1}) is much smaller than the phonon contribution ($\varepsilon_{phon}$). As a result, thermal expansion is typically analyzed by considering phonons alone, with minimal loss of accuracy. This provides the foundation for understanding the anisotropic thermal expansion in ErPd$_2$Si$_2$.

The chosen temperature range for our structural study was designed to minimize potential magnetostrictive effects \cite{li2012possible}. This ensures that the observed anisotropic thermal expansion in ErPd$_2$Si$_2$ primarily reflects the combined effects of lattice vibrations and electronic contributions. The expansion of the lattice constant \emph{a}, representing the sides of the tetragonal unit cell in the basal plane, is attributed to increased thermal energy, which enhances the vibrational amplitudes of atoms. This increased vibrational energy leads to a net expansion of the unit cell along the \emph{a} direction. At higher temperatures, electrons gain more kinetic energy, further promoting lattice expansion.

Thermal expansion fundamentally originates from atomic vibrations within the crystal lattice, which can be described through phonons. Phonons represent the quantized modes of lattice vibrations, reflecting the collective vibrational behavior of atoms in the lattice. As temperature rises, the excitation energy of phonons increases, resulting in more intense lattice vibrations. Due to the anharmonicity of the potential energy associated with lattice vibrations, the potential energy curve becomes asymmetric when atoms deviate from their equilibrium positions (e.g., steeper in the compression direction and flatter in the stretching direction). This asymmetry causes the average interatomic distance to increase as the amplitude of vibration grows, leading to positive thermal expansion.

Conversely, the contraction of the lattice constant \emph{c} can be attributed to an atypical form of the atomic potential. This refers to a potential energy curve with a shape or symmetry that deviates from typical materials, resulting in a reduction of interatomic distances as temperature increases. Furthermore, along the crystallographic \emph{c} axis, specific low-frequency phonon modes (soft phonon modes) or transverse phonon modes may dominate the lattice vibrations. These modes contribute to the observed decrease in the lattice constant \emph{c} with increasing temperature.

\subsection{Heat capacity}

Heat capacity measurements yield valuable insights into the thermodynamic properties of materials undergoing various phase transitions. During such transitions, a notable change in the material's internal energy occurs without a corresponding change in temperature. Figure~\ref{HC}a illustrates the heat capacity of a single crystal of ErPd$_2$Si$_2$ plotted against temperature. The heat capacity displays a sharp change, characterized by a $\lambda$-shape, with a maximum occurring at approximately 3.6 K, indicative of a phase transition. In this study, we define the transition temperature as the point at which a sudden change in the slope of the heat capacity occurs. This definition has been validated to be in accordance with magnetization studies \cite{PhysRevMaterials.4.094409}. Accordingly, $T_\textrm{N} \sim 4.20$ K was identified as the AFM phase transition temperature. Interestingly, two more anomalies were observed in the measured heat capacity: one at $T_1 \sim 3.0$ K (inset of Fig.~\ref{HC}a) and another at $T_2 \sim 15.31$ K (Fig.~\ref{HC}a). These observations suggest complex underlying mechanisms, prompting further investigation.

Heat capacity serves as a crucial indicator of changes in the arrangement of building components, including charges, spins, lattice, and orbital configurations, within a material. Consequently, it is intriguing to elucidate the physical mechanisms underlying the observed phase transitions in ErPd$_2$Si$_2$. Our previous research reported polarized and unpolarized neutron scattering studies on ErPd$_2$Si$_2$ \cite{Li2015-2}, revealing two AFM modulations with propagation wave vectors at \textbf{Q}$_{\pm}$ = ($H \pm 0.557(1)$, 0, $L \pm 0.150(1)$) and \textbf{Q}$_\texttt{C}$ = ($H \pm 0.564(1)$, 0, $L$), where $\left(H, L\right)$ are permissible Miller indices. Through comprehensive temperature- and magnetic field-dependent neutron scattering analyses, we associated the \textbf{Q}$_{\pm}$ modulation with localized 4\emph{f} moments emerging at $T_\textrm{N} \sim$ 4.20 K, while the \textbf{Q}$_\texttt{C}$ modulation corresponded well with itinerant spin-density waves, manifesting at $T_1$ and peaking at $\sim$ 3.5 K before rapidly diminishing at $T_\textrm{N}$ upon heating. A distinct anomaly, characterized by a sharp decrease in the first derivative of the heat capacity with respect to temperature, is observed at $T_\textrm{N} \sim 4.20$ K (inset of Fig.~\ref{HC}a). This anomaly corresponds to the AFM phase transition and links two phenomena upon cooling: the emergence of localized $4f$ moments of Er element and the presence of spin-density waves stemming from itinerant conduction bands. As the temperature decreases, both configurations become increasingly robust. Notably, the spin-density waves reach their maximum intensity at the heat-capacity peak before vanishing swiftly at $T_1$. Accordingly, the anomaly observed at $T_1 \sim 3.0$ K in heat capacity was attributed to the disappearance of spin-density waves. 

Conversely, no corresponding observations in following magnetization along the $c$-axis and resistance studies were made regarding the formation of the heat-capacity anomaly observed at $T_2 \sim 15.31$ K. A notable distinction is observed for ErPd$_2$Si$_2$ wherein the anomaly noted at $T_2 \sim 15.31$ K does not manifest in the heat capacity of GdPd$_2$Si$_2$ \cite{chatterjee2023unusual}. Previous magnetization studies conducted along the $<$1 1 0$>$ direction of ErPd$_2$Si$_2$ revealed a broad shoulder within the temperature range of 8--20 K, attributed to the presence of magnetic correlations above 5 K \cite{sampathkumaran2008magnetic}. However, our neutron diffraction analysis did not detect the diffusive magnetic signal associated with such short-range spin correlations above 6 K \cite{Li2015-2}. Previous heat capacity measurements \cite{cao2014single} show that ErPd$_2$Si$_2$ undergoes magnetic ordering at $T_\textrm{N} \approx$ 3.4 K. The broad hump observed around 15--20 K in magnetic entropy is attributed to the temperature-driven population of crystal-field split states \cite{cao2014single}. 

\subsection{Temperature dependent magnetization}

\begin{figure} [t]
\centering
\includegraphics[width = 0.82\textwidth] {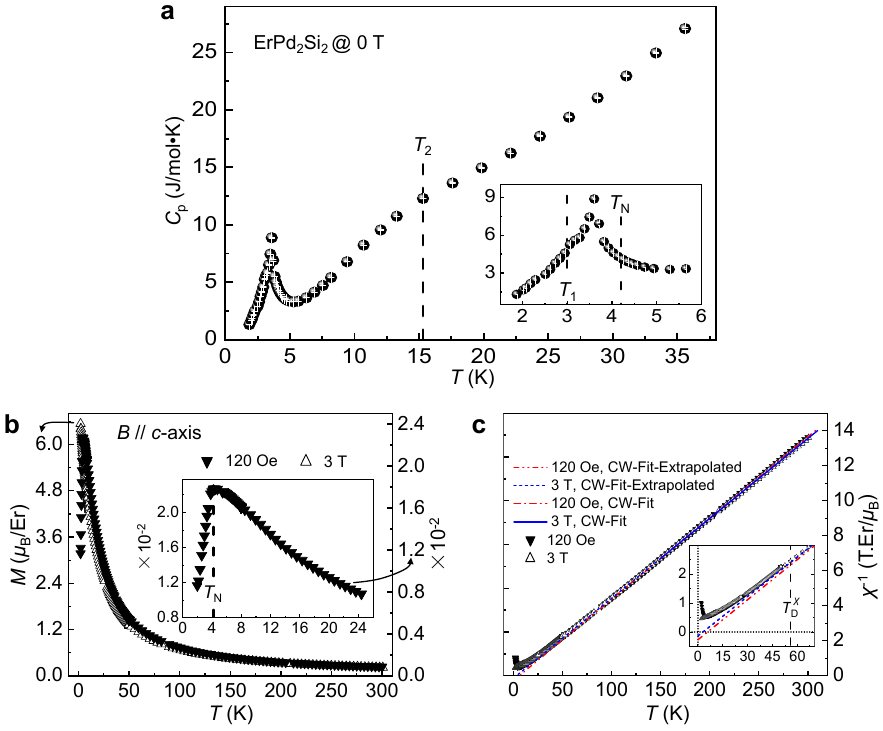}
\caption{\textbf{a} Heat capacity of an ErPd$_2$Si$_2$ single crystal as a function of temperature, measured under a 0 T magnetic field. The insert displays the data within the temperature range of 1.5--6 K. Three temperature points, $T_1 \sim 3.0$ K, $T_\textrm{N} \sim 4.20$ K, and $T_2 \sim 15.31$ K, have been identified as key indicators of phase transitions, as discussed in the text. The error bars represent estimated standard deviations derived from measurements.
\textbf{b} Measured magnetization (\emph{M}) as a function of temperature along the \emph{c} axis of an ErPd$_2$Si$_2$ single crystal, recorded at 120 Oe (downward filled triangles, right axis) and 3 T (upward open triangles, left axis). The inset illustrates the data within the temperature range of 0--26 K, where $T_\textrm{N} \sim 4.20$ K indicates a magnetic phase transition.
\textbf{c} Inverse magnetic susceptibility ($\chi^{-1}$) versus temperature, corresponding to \textbf{b}. The dash-dotted (120 Oe) and solid (3 T) lines denote the CW behavior of the data, described by Eq. (\ref{CW}). Both fits are extrapolated to $\chi^{-1} = 0$ (dashed-dot-dotted line, 120 Oe; short dashed line, 3 T) to reveal the PM Curie temperature ($\theta_{\textrm{CW}}$). The inset indicates $T_\textrm{D}^\chi \sim$ 55.8 K as the temperature point below which the CW fits deviate noticeably from the observed data.}
\label{HC}
\end{figure}

Temperature-dependent magnetization measurements of ErPd$_2$Si$_2$ were conducted with an applied magnetic field aligned parallel to the \emph{c} axis. As depicted in Fig.~\ref{HC}b, the ZFC magnetization rises smoothly as the temperature decreases until approximately 50 K, followed by a pronounced increase for $B$ = 3 T. However, the magnetization with $B$ = 120 Oe peaks at $T_\textrm{N} \sim$ 4.20 K (inset of Fig.~\ref{HC}b), consistent with our heat capacity measurements, before declining rapidly. No magnetization anomaly was observed at $T_1 \sim$ 3.0 K, suggesting that the spin-density waves disappearing at $T_1 \sim$ 3.0 K contribute minimally to the macroscopic magnetization. The magnetic anomaly indicative of the AFM phase transition only appears in the low-field ($B$ = 120 Oe) magnetization data at $T_\textrm{N} \sim$ 4.20 K, whereas the application of a high magnetic field ($B$ = 3 T) completely suppresses the magnetic anomaly, resulting in a continuous enhancement of magnetization rather than a peak. At 2 K, the measured magnetization $M$ = 6.54(1) $\mu_\textrm{B}$/Er at 3 T, and 0.0117(1) $\mu_\textrm{B}$/Er at 120 Oe. These results highlight the complex behavior of magnetization under varying magnetic fields.

In ErPd$_2$Si$_2$, the 4$f$ moments of Er are typically localized. The spin interactions between these 4$f$ moments are indirect and mediated through conduction bands. Previously, spin-density waves were also observed in ErPd$_2$Si$_2$, attributed to itinerant conduction bands. Both types of magnetism (localized and itinerant) exhibit complex cooperative behavior (3.5 K $< T < T_\textrm{N}$) and competitive behavior ($T_1 < T <$ 3.5 K) regarding temperature dependencies \cite{Li2015-2}, jointly contributing to the observed magnetization. These observations suggest a complex interplay between magnetic ordering, applied magnetic field, and temperature in ErPd$_2$Si$_2$.

The magnetization study revealed two notable features: (1) The measured magnetization at 2 K and 3 T, denoted as $M_{\textrm{(2K,3T)}}$, is 6.54(1) $\mu_\textrm{B}$/Er, approximately 37.61{\%} smaller than the theoretical saturation moment value (9 $\mu_\textrm{B}$) expected for Er$^{3+}$ ions. This observation suggests a localized characteristic of the measured moments. It is noteworthy that this comparison leans towards empiricism, as Er in ErPd$_2$Si$_2$ does not form pure ionic (3+) ions; instead, Er, Pd, and Si ions share conduction electrons. (2) Additionally, the application of a 3 T magnetic field eliminates the evidence of long-range ordered and localized 4$f$ moments. This suppression may be attributed to: (i) The applied magnetic field of 3 T may exert a strong Zeeman effect on the magnetic moments of the Er ions, destabilizing the AFM ordering. (ii) The energy level associated with the applied magnetic field of 3 T may become comparable to or larger than the energy associated with the AFM ordering, leading to the suppression of the ordered state. Consequently, the application of a 3 T magnetic field readily reorients the localized AFM spins through spin-flop and spin-flip transitions \cite{sampathkumaran2008magnetic, li2016possible}. The suppression of $T_\textrm{N}$ (the temperature at which the AFM ordered ground state is suppressed) by a high magnetic field has been previously reported in CePd$_2$Si$_2$ \cite{sheikin2002specific} and YbRh$_2$Si$_2$ \cite{custers2001low}.

In a pure paramagnetic (PM) state, the inverse magnetic susceptibility, denoted as $\chi^{-1} = B/M$, typically increases linearly with temperature, conforming well to the Curie-Weiss (CW) law,
\begin{eqnarray}
\chi^{-1}(T) = \frac{3k_\textrm{B}(T - \theta_{\textrm{CW}})}{N_\textrm{A} \mu^2_{\textrm{eff}}},
\label{CW}
\end{eqnarray}
where $k_\textrm{B}$ = 1.38062 $\times$ 10$^{-23}$ J/K represents the Boltzmann constant, $\theta_{\textrm{CW}}$ denotes the PM Curie temperature, $N_\textrm{A}$ = 6.022 $\times$ 10$^{23}$ mol$^{-1}$ signifies the Avogadro's number, and $\mu_{\textrm{eff}}$ = $g_J \mu_\textrm{B} \sqrt{J(J + 1})$ stands for the effective PM moment. We computationally derived the theoretical value of $\mu_{\textrm{eff}}$ for ErPd$_2$Si$_2$, considering the quantum numbers of Er$^{3+}$ ions (4$f^{11}$, $S = 3/2$, $L = 6$, $J = 15/2$, and $g_J = 1.2$), yielding $\mu_{\textrm{eff{\_}theo}} = 9.581$ $\mu_\textrm{B}$. As depicted in Fig.~\ref{HC}c, the magnetization data acquired at 120 Oe and 3 T within the temperature range of 150--300 K were fitted using Eq.~\ref{CW}, displaying excellent adherence to CW behavior. These fittings yielded effective PM moments, denoted as $\mu_{\textrm{eff{\_}meas}}$, of 9.808(3) $\mu_\textrm{B}$ at 120 Oe and 9.889(3) $\mu_\textrm{B}$ at 3 T. Both measured $\mu_{\textrm{eff{\_}meas}}$ values exhibited minimal discrepancy and closely matched the expected theoretical value of $\mu_{\textrm{eff{\_}theo}} = 9.581$ $\mu_\textrm{B}$, indicating the dominant magnetic contribution from Er in ErPd$_2$Si$_2$. Although the application of a 3 T magnetic field can suppress the AFM ordered spin state, it has negligible effect on the PM state. This underscores that ErPd$_2$Si$_2$ indeed remains in a pure PM state above 150 K. As shown in the inset of Fig.~\ref{HC}c, we are thus more confident in concluding that below $T_\textrm{D}^\chi \sim$ 55.8 K, the measured magnetization data deviate from the CW law, exhibiting noticeable upward deviations from theoretical predictions. From these analyses, we therefor infer that the persistence of AFM spin correlations from low temperatures up to $T_\textrm{D}^\chi \sim$ 55.8 K, significantly surpassing $T_\textrm{N} \sim$ 4.20 K. The presence of short-range AFM interactions may be another reason for the observed heat capacity anomaly at $T_2 \sim 15.31$ K and the broad magnetic anomaly observed at 8--20 K \cite{sampathkumaran2008magnetic}. Corresponding neutron diffusive information was not identified \cite{Li2015-2}, likely due to the weak intensity of scattered neutrons.

The CW temperature provides valuable insight into the underlying spin interactions within a magnetic material. By extrapolating the CW fits across the entire temperature range studied (Fig.~\ref{HC}c), we determined the CW temperature to be $\theta^{\textrm{120Oe}}_{\textrm{CW}} =$ 5.8(1) K at 120 Oe and $\theta^{\textrm{3T}}_{\textrm{CW}} =$ 3.0(1) K at 3 T. Both $\theta_{\textrm{CW}}$ values are positive, indicating the existence of ferromagnetic spin interactions along the crystallographic \emph{c}-axis. The application of a 3 T magnetic field resulted in a reduction of the CW temperature from 5.8 to 3.0 K, representing a decrease of $\sim$ 48.28{\%}. This significant reduction suggests a weakening of the ferromagnetic interactions and a potential enhancement of AFM interactions along the crystallographic \emph{c}-axis.

According to previous studies on the applied magnetic-field dependent magnetization of ErPd$_2$Si$_2$ \cite{sampathkumaran2008magnetic, uchima2018magnetic}, the magnetic moments along the $c$-axis reach near saturation at 3 T and 1.8 K, indicating a flopped and flipped state of the AFM spins. In contrast, the magnetic moments remain far below saturation at 6 T when the applied magnetic field $B$ is aligned along the $\langle$1 1 0$\rangle$ direction. These observations confirm the $c$-axis as the AFM easy axis, while the $\langle$1 1 0$\rangle$ orientation represents the hard direction.

\subsection{Temperature dependent resistance}

The electrical resistivity of single-crystal ErPd$_2$Si$_2$ was previously investigated with excitation currents applied along either the $<$0 0 1$>$ or the $<$1 1 0$>$ direction \cite{sampathkumaran2008magnetic}. Two primary observations were made: (i) The emergence of magnetic ordering at low temperatures exhibited no discernible effect on the electrical resistivity \cite{sampathkumaran2008magnetic}, consistent with findings from our previous study \cite{Li2015-2}. (ii) Application of magnetic fields up to 14 T did not induce changes in either the values of electrical resistivity or the features of temperature-dependent electrical resistivity \cite{sampathkumaran2008magnetic}. 

\begin{figure} [t]
\centering
\includegraphics[width = 0.82\textwidth] {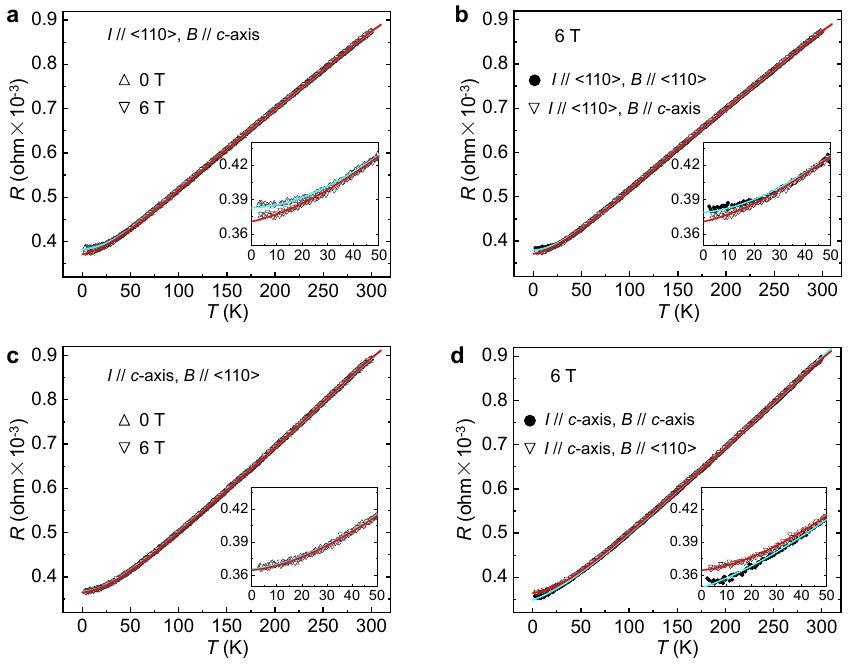}
\caption{Measured resistance, $R$, of an ErPd$_2$Si$_2$ single crystal as a function of temperature.
\textbf{a} $R$ values were obtained with current ($I$) flowing along the $\langle$1 1 0$\rangle$ direction, and the magnetic field ($B$) (upward open triangles, 0 T; downward open triangles, 6 T) aligned parallel to the $c$ axis.
\textbf{b} $R$ values were measured with current $I$ flowing along the $\langle$1 1 0$\rangle$ direction, and the magnetic field $B$ either parallel to the $\langle$1 1 0$\rangle$ direction (filled circles, 6 T) or along the $c$ axis (downward open triangles, 6 T).
\textbf{c} $R$ values were measured with current $I$ flowing along the $c$ axis, and the magnetic field $B$ (upward open triangles, 0 T; downward open triangles, 6 T) parallel to the $\langle$1 1 0$\rangle$ direction.
\textbf{d} $R$ values were measured with current $I$ flowing along the $c$ axis, and the magnetic field $B$ either parallel to the $c$ axis (filled circles, 6 T) or applied along the $\langle$1 1 0$\rangle$ direction (downward open triangles, 6 T).
In panels \textbf{a}-\textbf{d}, solid lines represent fits to the data using Eq. (\ref{RTfit}) (see details in the text), and insets provide enlarged images of the data within the temperature range of 0--50 K.}
\label{RT}
\end{figure}

In this study, we conduced a more comprehensive investigation of the electrical resistance of an ErPd$_2$Si$_2$ single crystal (Fig.~\ref{RT}) and identified some previously overlooked interesting characteristics. Figure~\ref{RT} presents the measured temperature-dependent resistance values obtained along multiple crystallographic directions under applied magnetic fields of 0 and 6 T. As the temperature decreases from room temperature, the electrical resistance, denoted as $R$, measured with zero applied magnetic field and an excitation current, $I$, along the $<$1 1 0$>$ direction, exhibits a smooth decrease (Fig.~\ref{RT}a). This trend persists even across the observed phase transitions, as evidenced by the heat capacity anomalies (Fig.~\ref{HC}a) and magnetization data (Fig.~\ref{HC}b), in agreement with prior studies \cite{sampathkumaran2008magnetic, Li2015-2}. Notably, a comparison reveals significant differences in resistance values at low temperatures when the applied magnetic field, $B$, is increased along the $c$-axis, reaching 6 T (inset of Fig.~\ref{RT}a). Lower resistance values are observed under stronger magnetic fields, while at high temperatures, the resistance values coincide. Furthermore, when the applied magnetic field ($B =$ 6 T) changes orientation from the $c$-axis to the $<$1 1 0$>$ direction, where $I \parallel B$, an increase in resistance values at low temperatures is observed (inset of Fig.~\ref{RT}b), while the high-temperature $R$ values remain consistent. Based on these observations, we conclude that when $I \parallel $ $<$1 1 0$>$: (i) $R$ values show no response to magnetic phase transitions. (ii) A magnetoresistance effect is observed at low temperatures when an applied magnetic field of $B =$ 6 T is either parallel or perpendicular to the current $I$.

We further investigated the case where $I \parallel c$-axis, as depicted in Fig.~\ref{RT}c, d. The resistance still exhibits a smooth increase upon warming (Fig.~\ref{RT}c), and no obvious magnetoresistance effect is observed when $B$ (= 6 T) is oriented along the $<$1 1 0$>$ direction (i.e., $B$ perpendicular to the current $I$ orientation) (inset of Fig.~\ref{RT}c). When the applied magnetic field ($B =$ 6 T) changes orientation from the $<$1 1 0$>$ direction to the $c$-axis, i.e., $B \parallel I$, the resistance values at low temperatures decrease (inset of Fig.~\ref{RT}d), while the high-temperature $R$ values still remain consistent. Therefore, We conclude that when $I \parallel c$-axis: (i) $R$ values also exhibit no response to magnetic phase transitions when $B =$ 0 T or 6 T is perpendicular to the current (Fig.~\ref{RT}c). (ii) Notably, the measured resistance values increase below $T_\textrm{N}$ when both $B$ (= 6 T) and $I \parallel c$-axis (inset of Fig.~\ref{RT}d). (iii) A magnetoresistance effect is observed at low temperatures between the applied magnetic field of $B =$ 6 T alongside and perpendicular to the current $I$ (inset of Fig.~\ref{RT}d). These findings provide valuable insights into the magnetotransport behavior of ErPd$_2$Si$_2$ and its response to magnetic field variations along different crystallographic orientations.

We previously established that the moments in the magnetically ordered state of ErPd$_2$Si$_2$ lie in the ($H$, $L$) scattering plane, with the AFM axis aligned parallel to the $c$-axis \cite{Li2015-2}. In our temperature-dependent magnetization study (Fig.~\ref{HC}b), we observed the destruction of the magnetically ordered state under an applied magnetic filed of $B =$ 3 T along the $c$-axis. We attribute this phenomenon to possible spin-flop and spin-flip transitions \cite{li2016possible}. These applied magnetic-field-driven magnetic phase transitions also result in an increase in resistance (inset of Fig.~\ref{RT}d).

\subsection{Temperature dependent magnetoresistance effect}

\begin{figure} [t]
\centering
\includegraphics[width = 0.48\textwidth] {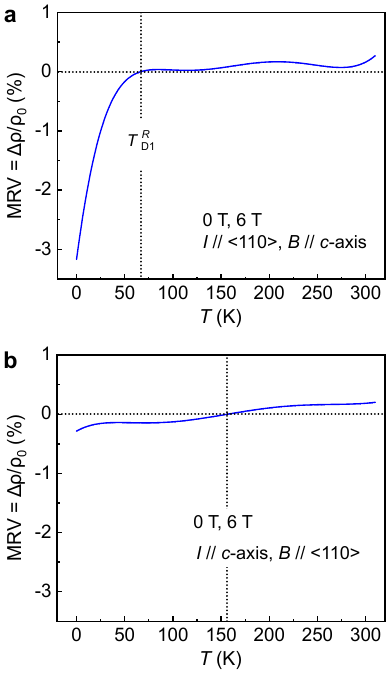}
\caption{Calculated MRVs of an ErPd$_2$Si$_2$ single crystal as a function of temperature.
\textbf{a} The MRVs were calculated in the magnetic field range of 0 to 6 T, with the current, $I$, flowing along the $\langle$1 1 0$\rangle$ direction, and the magnetic field, $B$, aligned parallel to the $c$ axis.
\textbf{b} The MRVs were calculated with the current, $I$, flowing along the $c$ axis, and the magnetic field, $B$, aligned parallel to the $\langle$1 1 0$\rangle$ direction, spanning a field range of 0 to 6 T.
In panels \textbf{a}, $T_\textrm{D1}^R$ denotes the temperature at which the MRV undergoes significant changes.}
\label{MRVs}
\end{figure}

The temperature-dependent change in electrical resistivity of LuPdSi$_3$, driven by electron-phonon interactions, was analyzed using the Bloch-Gr$\ddot{\textrm{u}}$neisen law \cite{Cao2013-1}. However, the electrical resistivity deviates from this law to some extent due to resistivity anisotropy. Therefore, to quantitatively analyze the temperature-dependent resistance values of ErPd$_2$Si$_2$ as depicted in Fig.~\ref{RT}, we performed a schematic fitting of the data to the equation:
\begin{eqnarray}
R(T) = \textrm{C} + \textrm{N}T + \textrm{O}T^2 + \textrm{P}T^3 + \textrm{Q}T^4 + \textrm{S}T^5,
\label{RTfit}
\end{eqnarray}
where C represents the residual resistance, N, O, P, Q, and S are constants for the quintic function. The solid lines in Fig.~\ref{RT} represent the fittings, which agree well with the resistance measurements. We define the magnetoresistance value (MRV) as:
\begin{eqnarray}
\textrm{MRV}(B, T, \gamma, \theta)~=~\frac{R(B, T, \gamma, \theta) - R(0, T, \gamma, \theta)}{R(0, T, \gamma, \theta)} \times 100\%,
\label{MRV}
\end{eqnarray}
where $R(B, T, \gamma, \theta)$ and $R(0, T, \gamma, \theta)$ represent the measured resistances with and without an applied magnetic field ($B$), respectively. These resistances are determined at a specified temperature (\emph{T}) and angles ($\gamma$ and $\theta$, as illustrated in Figs.~\ref{Rrota2}a and \ref{Rrota3}a). The calculated MRVs versus temperature are shown in Fig.~\ref{MRVs} using Equ.~(\ref{MRV}).

When $I$ is parallel to $<$1 1 0$>$, and $B$ aligns with the $c$-axis, varying from 0 to 6 T, the MRVs remain positive until reaching $T_\textrm{D1}^R =$ 66.5 K upon cooling. At this point, a transition from positive to negative MRVs occurs, with the smallest extracted negative MRV recorded at -3.17\% (Fig.~\ref{MRVs}a). It's worth noting that the observed $T_\textrm{D1}^R =$ 66.5 K coincides with $T_\textrm{D}^\chi \sim$ 55.8 K, where the measured magnetization data deviate from the CW law (Fig.~\ref{HC}c). Therefore, we infer the following: (i) Above $T_\textrm{D1}^R$, ErPd$_2$Si$_2$ remains in a pure PM state, where the application of a magnetic field increases the resistance compared to the zero-field condition, leading to positive MRVs. (ii) Below $T_\textrm{D1}^R$, the emergence of AFM correlations results in negative MRVs, where the application of a magnetic field decreases the electrical resistance. When the direction of the current $I$ aligns with the crystallographic $c$-axis and the magnetic field $B$ aligns with the $<$1 1 0$>$ direction, the absolute maximum MRV decreases significantly, by approximately 91\%, indicating the absence of a magnetoresistance effect. This is likely attributed to the single-ion anisotropy, which dictates the preferred orientation of spin moments along the \emph{c}-axis rather than within the basal \emph{ab}-plane. However, it is noteworthy that a sign transition from positive to negative occurs at $\sim$ 156 K, as illustrated in Fig.~\ref{MRVs}b. 

\begin{figure} [t]
\centering
\includegraphics[width = 0.48\textwidth] {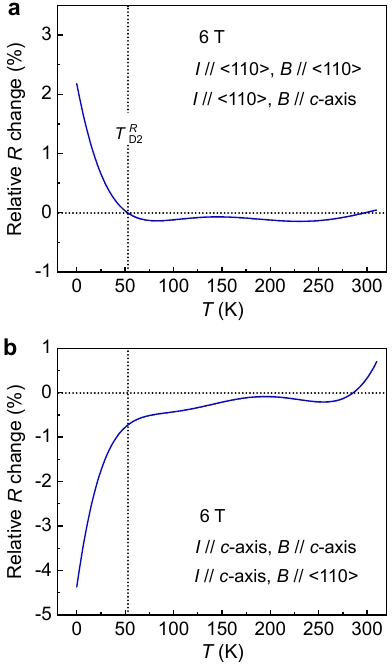}
\caption{Relative resistance change of an ErPd$_2$Si$_2$ single crystal as a function of the applied magnetic field direction.
\textbf{a} The relative resistance change was measured with the current, $I$, flowing along the $\langle$1 1 0$\rangle$ direction, and the magnetic field, $B$, set at 6 T, rotating between the $\langle$1 1 0$\rangle$ and $c$-axis directions.
\textbf{b} The relative resistance change was measured with the current, $I$, flowing along the $c$ axis, and the magnetic field, $B$, set at 6 T, rotating between the $c$-axis and $\langle$1 1 0$\rangle$ directions.
In panels \textbf{a}, $T_\textrm{D2}^R$ denotes the temperature at which the relative resistance change undergoes significant variations.}
\label{RDirection}
\end{figure}

Interestingly, when the applied magnetic field direction is rotated from the $c$-axis to the $<$1 1 0$>$ direction, the relative resistance change transitions from negative to positive at $T_\textrm{D2}^R =$ 52.5 K, reaching a maximum relative \emph{R} change of 2.18\% (Fig.~\ref{RDirection}a). In contrast, the lowest negative relative \emph{R} change (-4.36\%) is observed when the current ($I$) is parallel to the $c$-axis, and the applied magnetic field ($B =$ 6 T) rotates from the $<$1 1 0$>$ direction to the $c$-axis (Fig.~\ref{RDirection}b). 

These observations indicate that the measured magnetoresistance effect exhibits strong anisotropy. Specifically, variations in either the strength or the direction of the applied magnetic field lead to significant changes in the magnetoresistance.

\subsection{Angular dependent resistance}

\begin{figure} [t]
\centering
\includegraphics[width = 0.82\textwidth] {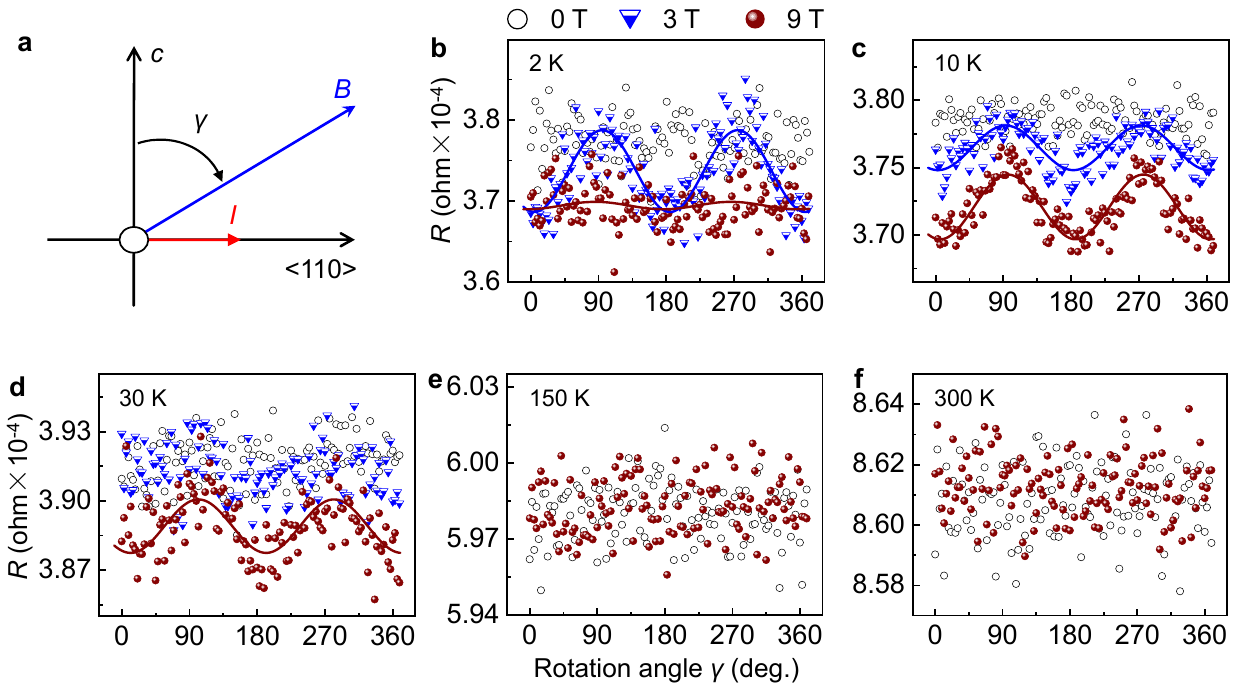}
\caption{Angular $\gamma$ dependence of resistance $R$ in an ErPd$_2$Si$_2$ single crystal under applied magnetic fields of 0 T (open circles), 3 T (downward half-filled triangles), and 9 T (filled circles). In \textbf{a}, for the resistance measurements, current $I$ is directed along the $\langle$1 1 0$\rangle$ direction, while the applied magnetic field rotates away from the \emph{c} axis with an angle $\gamma$. The resistance values were measured at temperatures of \textbf{b} 2 K, \textbf{c} 10 K, \textbf{d} 30 K, \textbf{e} 150 K, and \textbf{f} 300 K. The solid lines in \textbf{b}-\textbf{d} represent fits of the data using a combined sine-cosine function.}
\label{Rrota2}
\end{figure}

To explore the intricate interplay between magnetic and electronic phenomena in ErPd$_2$Si$_2$, we investigated the sensitivity of resistance to variation in magnetic field direction. As depicted in Fig.~\ref{Rrota2}a, the current $I$ is directed along the $\langle$1 1 0$\rangle$ direction, while the magnetic field $B$ rotates from the $c$-axis to the $\langle$1 1 0$\rangle$ direction. The angle $\gamma$ between the $c$-axis and the direction of $B$ ranges from 0 to 360$^{\circ}$ for one complete rotation with a step size of 3$^{\circ}$. Measurements were conducted under magnetic fields of 0 T, 3 T, and 9 T, respectively. These initial observations set the stage for a detailed analysis of anisotropic resistance behavior.

In Fig.~\ref{Rrota2}b-d, resistance values under 0 T exhibited no anisotropic behavior at temperatures of 2 K, 10 K, and 30 K, remaining relatively higher values. Similarly, in Fig.~\ref{Rrota2}e-f, resistance values under magnetic fields ranging from 0 to 9 T showed no appreciable differences within the present experimental accuracy at temperatures of 150 K and 300 K, indicating a lack of anisotropic behavior across the applied magnetic field strengths and the studied temperature range. These findings suggest that the anisotropic behavior of resistance is temperature and magnetic field dependent.

At 3 T, a twofold symmetry emerged at 2 K (Fig.~\ref{Rrota2}b) and 10 K (Fig.~\ref{Rrota2}c), gradually diminishing as the temperature rose to 30 K (Fig.~\ref{Rrota2}d). Within this symmetry, the smallest $R$ values occurred at 0$^{\circ}$ and 180$^{\circ}$, corresponding to $B \parallel c$-axis, while the largest $R$ values occurred at 90$^{\circ}$ and 270$^{\circ}$, corresponding to $B \parallel$ $\langle$1 1 0$\rangle$. During the rotation of $B$ from 0 to 90$^{\circ}$, the magnetic moments of Er decreased along the $c$-axis and increased in the $\langle$1 1 0$\rangle$ direction. When $B$ was parallel to the easy magnetic axis $c$, the magnetic moments of Er were aligned by the field towards the $c$-axis.

At 9 T, the twofold symmetry was rarely observed at 2 K, and the resistance remained low. As temperature increased, the twofold symmetry of resistance reappeared at 10 K and disappeared above 30 K. Resistance at 9 T was consistently lower than at 0 T and 3 T (Fig.~\ref{Rrota2}b-d) across all rotation angles $\gamma$. Only when the temperature exceeded the Debye temperature did the differences in resistance under each magnetic field vanish, along with the twofold symmetry of the angular-dependent resistances (Fig.~\ref{Rrota2}e-f). These observations provide a foundation for further investigation of the twofold symmetric resistance characteristics in ErPd$_2$Si$_2$.

\begin{figure} [t]
\centering
\includegraphics[width = 0.82\textwidth] {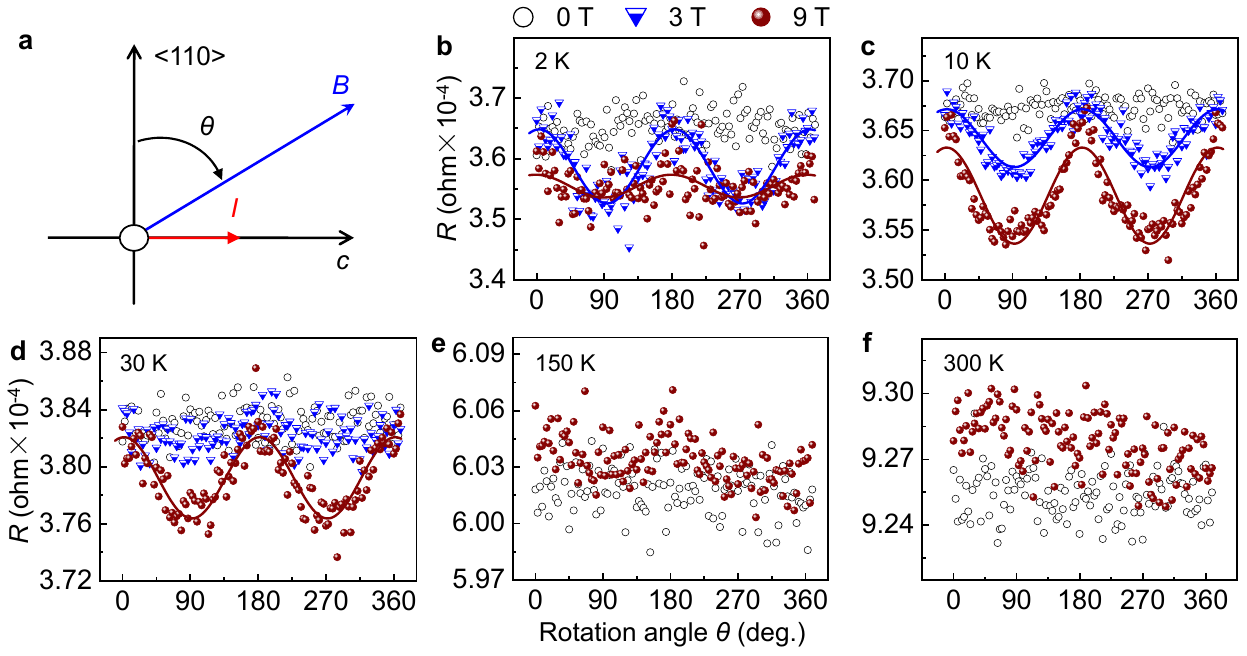}
\caption{Angular $\theta$-dependent resistance $R$ of an ErPd$_2$Si$_2$ single crystal under applied magnetic fields of 0 T (open circles), 3 T (downward half-filled triangles), and 9 T (filled circles). In \textbf{a}, for the resistance measurements, current $I$ is applied along the \emph{c} axis, and the applied magnetic field rotates with an angle $\theta$ deviating from the $\langle$1 1 0$\rangle$ direction. The resistance values were measured at temperatures of \textbf{b} 2 K, \textbf{c} 10 K, \textbf{d} 30 K, \textbf{e} 150 K, and \textbf{f} 300 K. The solid lines in \textbf{b}-\textbf{d} represent fits of the data with a combined sine-cosine function.}
\label{Rrota3}
\end{figure}

We further investigated the observed characteristic of twofold symmetric resistance in ErPd$_2$Si$_2$ when $I$ is parallel to the $c$-axis and $B$ rotates away from the $\langle$1 1 0$\rangle$ direction with an angle $\theta$ (0-360$^\circ$), as displayed in Fig.~\ref{Rrota3}. Similar to the results observed in Fig.~\ref{Rrota2}, the zero-field resistance ($R_{\textrm{0T}}$) is higher than those at $B =$ 3 T and 9 T below 30 K (Fig.~\ref{Rrota3}b-d), whereas $R_{\textrm{0T}}$ is lower at 150 K (Fig.~\ref{Rrota3}e) and 300 K (Fig.~\ref{Rrota3}f). Importantly, $R_{\textrm{0T}}$ values show no dependence on angle $\theta$ within the entire studied temperature range, as do $R_{\textrm{3T}}$ and $R_{\textrm{9T}}$ at 150 K and 300 K. Moreover, $R_{\textrm{3T}}$ and $R_{\textrm{9T}}$ display much clearer twofold symmetry (Fig.~\ref{Rrota3}b-d), except for $R_{\textrm{3T}}$ at 30 K (Fig.~\ref{Rrota3}d). Interestingly, $R_{\textrm{3T}}$ and $R_{\textrm{9T}}$ have the smallest values at $\theta =$ 90$^{\circ}$ and 270$^{\circ}$, whereas they reach maximum values at $\theta =$ 0$^{\circ}$ and 180$^{\circ}$. These findings suggest that the twofold symmetry is sensitive to both magnetic field strength and temperature.

It is noted that the twofold symmetry of resistance appears at 2 K (below $T_\textrm{N} \sim$ 4.20 K) and 10 K and 30 K (above $T_\textrm{N}$ but below $T_\textrm{D}^\chi \sim$ 55.8 K). Below $T_\textrm{N}$, the spin moments are aligned along the $c$-axis. Therefore, we infer that above $T_\textrm{N}$, the short-range spin fluctuations also display a preferred orientation (along the $c$-axis) due to single-ion anisotropy. It seems that above $T_\textrm{N}$, the twofold symmetry is much easier to establish at 9 T due to much weaker spin interactions compared to the magnetically ordered state. Therefore, the twofold symmetry observed in the electric property of ErPd$_2$Si$_2$ is strongly correlated with the spin-moment configuration rather than spin-interaction strength. These observations sharply contrast with those observed in superconductors, where the behavior of angular-dependent magnetoresistance is suppressed by a sufficiently strong magnetic field \cite{ZhenZhao77406}.

\subsection{Applied magnetic field dependent resistance}

While examining the angular-dependent resistance, distinct resistance values are observed at 0 T, 3 T, and 9 T below 30 K. Subsequently, we delve into the detailed investigation of the applied magnetic field-dependent resistance of the ErPd$_2$Si$_2$ single crystal.

\begin{figure} [t]
\centering
\includegraphics[width = 0.82\textwidth] {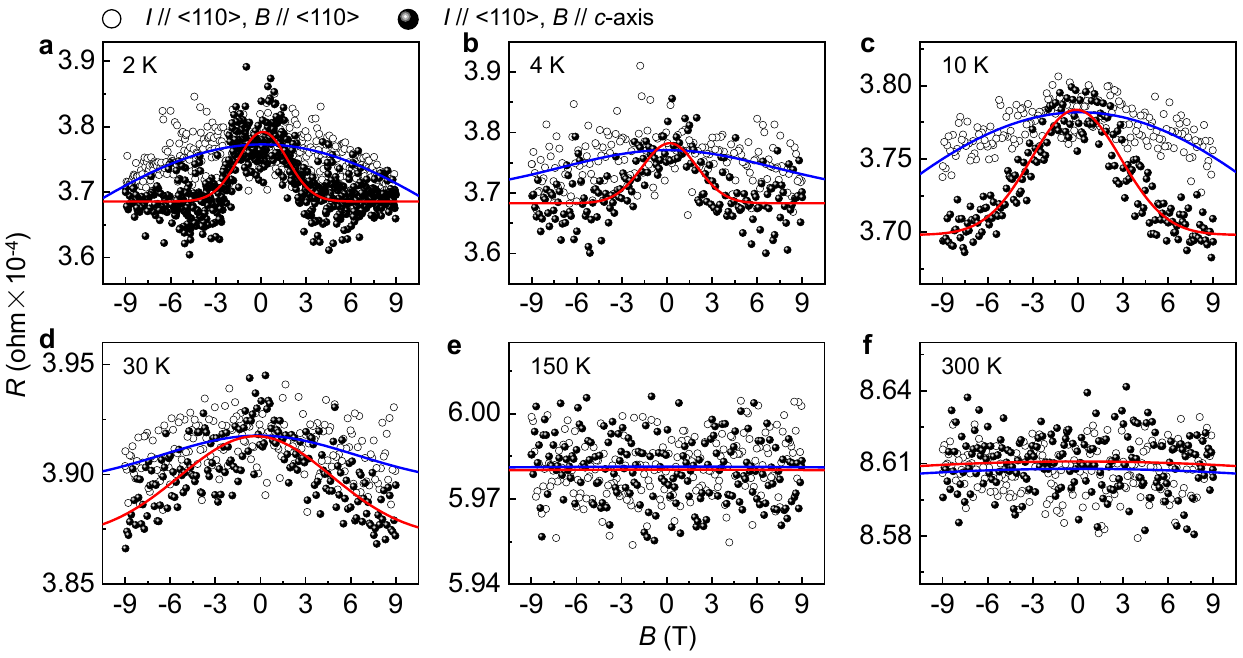}
\caption{Magnetic field $B$-dependent resistance $R$ of an ErPd$_2$Si$_2$ single crystal. The resistance values were measured at temperatures of \textbf{a} 2 K, \textbf{b} 4 K, \textbf{c} 10 K, \textbf{d} 30 K, \textbf{e} 150 K, and \textbf{f} 300 K. In measurements \textbf{a}-\textbf{f}, the current $I$ was applied along the $\langle$1 1 0$\rangle$ direction, and the applied magnetic field was oriented either along the $\langle$1 1 0$\rangle$ direction (open circles) or parallel to the $c$ axis (filled circles). The solid lines in \textbf{a}-\textbf{f} represent fits of the data with a Gaussian function.}
\label{RB2}
\end{figure}

As depicted in Fig.~\ref{RB2}, the current $I$ is directed along the $\langle$1 1 0$\rangle$ direction, while magnetic fields $B$ are applied along either the $\langle$1 1 0$\rangle$ or the $c$-axis. The magnetic field strength varies from -9 to +9 T within the temperature range of 2--300 K. It is evident that the resistance values when $B$ is parallel to $\langle$1 1 0$\rangle$ are greater than those when $B$ is parallel to the $c$-axis, consistent with our investigation of angular $\gamma$-dependent resistance. When $B$ is parallel to $I$, $R_{\langle110\rangle}$ values exhibit weak magnetic field dependence below 30 K. As $B$ increases from 0 T to 9 T, the $R_{\langle110\rangle}$ value decreases slightly, albeit this quadratic-like trend is less pronounced (Fig.~\ref{RB2}a-d). In contrast, when $B$ is perpendicular to $I$, $R_{\langle110\rangle}$ demonstrates a clear magnetic field dependence. For instance, at 2 K, $R^{B \perp I}_{\langle110\rangle}$ values initially linearly decrease as $B$ increases from 0 to 3 T, after which they stabilize up to 9 T (Fig.~\ref{RB2}a). The magnetic field point ($B_{\textrm{LS}}$), corresponding to the $R$ transition from linear $B$ dependence to a stable state, increases to approximately 4.5 T (Fig.~\ref{RB2}b), 6 T (Fig.~\ref{RB2}c), and even higher than 9 T (Fig.~\ref{RB2}d). This magnetic field dependence of $R$ vanishes at 150 K (Fig.~\ref{RB2}e) and 300 K (Fig.~\ref{RB2}f).

\begin{figure} [t]
\centering
\includegraphics[width = 0.82\textwidth] {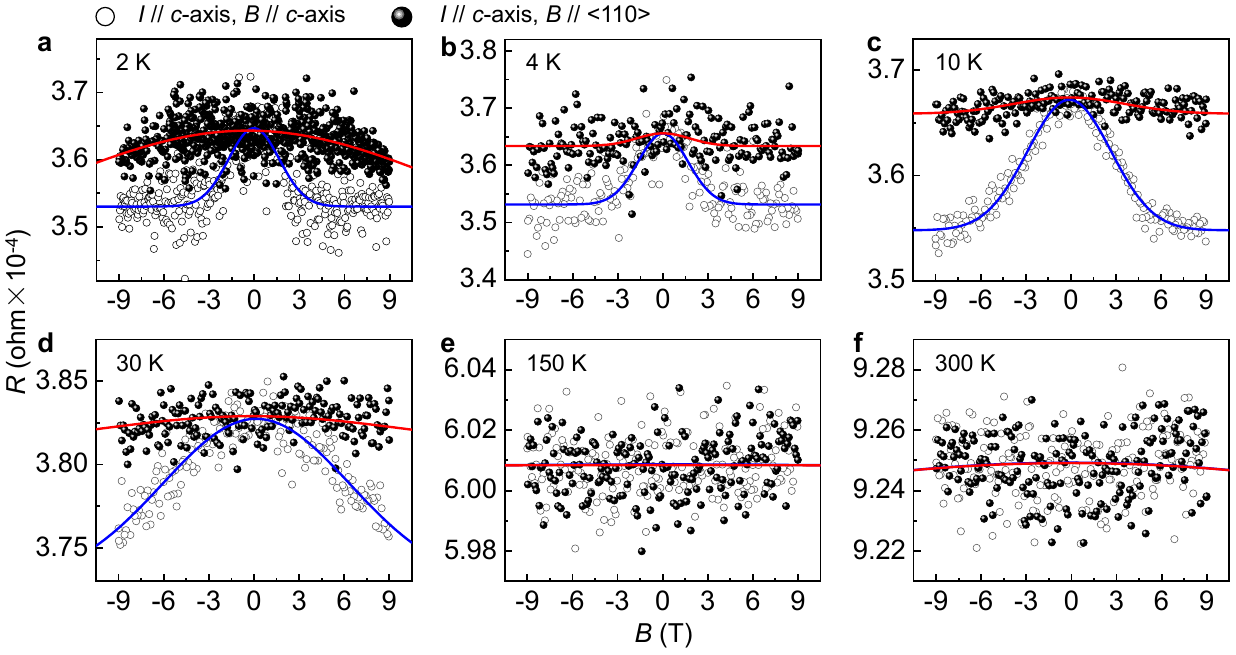}
\caption{Magnetic field $B$-dependent resistance $R$ of an ErPd$_2$Si$_2$ single crystal. The resistance values were measured at temperatures of \textbf{a} 2 K, \textbf{b} 4 K, \textbf{c} 10 K, \textbf{d} 30 K, \textbf{e} 150 K, and \textbf{f} 300 K. In measurements \textbf{a}-\textbf{f}, the current $I$ was applied along the $c$ axis, while the applied magnetic field was directed either along the $c$ axis (open circles) or parallel to the $\langle$1 1 0$\rangle$ direction (filled circles). The solid lines in \textbf{a}-\textbf{f} represent fits of the data using a Gaussian function.}
\label{RB3}
\end{figure}

The aforementioned observations are more prominent when the current $I$ is applied along the $c$-axis direction (Fig.~\ref{RB3}), where $R^{B \perp I}_{c{\textrm{-axis}}}$ $>$ $R^{B \parallel I}_{c{\textrm{-axis}}}$ below 30 K (Fig.~\ref{RB3}a-d). This aligns with the results from Fig.~\ref{Rrota3}. At 150 K (Fig.~\ref{RB3}e) and 300 K (Fig.~\ref{RB3}f), there is no magnetic field dependence of $R$.

\section{Conclusion}

In summary, our comprehensive investigation into the structure, heat capacity, magnetization, and magnetoresistance behaviors of single-crystal ErPd$_2$Si$_2$ highlights the intricate interplay between magnetic and electronic phenomena in this compound. The study makes several key observations:
(1) Opposite temperature-dependent changes in the lattice parameters of ErPd$_2$Si$_2$ were observed. Specifically, the lattice constant \emph{a} expands, while \emph{c} contracts between 15 K and 300 K. This results in a positive thermal expansion of the unit cell volume, providing new insights into the lattice dynamics of ErPd$_2$Si$_2$. The anisotropic thermal expansion is attributed to the combined effects of lattice vibrations and electronic contributions, which have not been previously reported for this compound. 
(2) Three distinct temperature regimes were identified through heat capacity measurements: 
(i) $T_1 \sim 3.0$ K: Associated with the disappearance of the SDW. 
(ii) $T_\textrm{N} \sim 4.20$ K: Marks the onset of an incommensurate AFM structure. 
(iii) $T_2 \sim 15.31$ K: Linked to crystal-field splitting and/or the presence of short-range spin fluctuations. These findings shed light on the interplay between localized 4\emph{f} moments and itinerant conduction electrons. 
(3) A unique twofold symmetry in angular-dependent resistance was observed at low temperatures (e.g., 2 K, 10 K, and 30 K) under applied magnetic fields (3 T and 9 T). This symmetry is attributed to the alignment of spin moments relative to the magnetic field, reflecting the sensitivity of ErPd$_2$Si$_2$ to magnetic field orientation. The MR behavior exhibits anisotropy, with resistance values varying significantly depending on the strength and direction of the applied magnetic field. 
(4) Magnetic field-driven spin reorientation was identified. The application of a high magnetic field (3 T) suppresses the AFM phase transition anomaly, likely due to spin-flop and spin-flip transitions. The study identifies the \emph{c}-axis as the AFM easy axis, while the $⟨1 1 0⟩$ direction represents the hard axis. This clarifies the magnetic anisotropy and spin configurations in ErPd$_2$Si$_2$. 
(5) Persistence of AFM correlations beyond the N$\acute{e}$el temperature: This persistence is evidenced by deviations from the CW law and is further linked to the heat capacity anomaly at $T_2 \approx$ 15.31 K. These results provide key insights into the relationship between AFM correlations and lattice dynamics. 
(6) Correlation between angular-dependent resistance and spin anisotropy: The twofold symmetry in resistance is strongly correlated with the spin-moment configuration, emphasizing the role of single-ion anisotropy. These findings build on the persistence of AFM correlations, highlighting the sensitivity of electronic transport to spin configurations. 
(7) Temperature-dependent magnetotransport behavior: A detailed analysis of the magnetic field-dependent resistance along different crystallographic directions was performed. These findings deepen the understanding of the interplay between magnetic field orientation, spin configurations, and electronic transport in ErPd$_2$Si$_2$, advancing knowledge of its magnetic and electronic properties.

\clearpage

\backmatter

\section*{Acknowledgments}

C.D. Cao and H.-F. Li express their gratitude to Wolfgang L$\ddot{\textrm{o}}$ser from the {Leibniz}-{Institut} für Festkörper{-} und Werkstoffforschung (IFW) Dresden, Germany, for pre-reviewing the manuscript and offering valuable suggestions. This work was supported by the Science and Technology Development Fund, Macao SAR (File Nos. 0090{/}2021{/}A2 and 0104{/}2024{/}AFJ) and University of Macau (MYRG{-}GRG2024{-}00158{-}IAPME). L.W. Li acknowledges the support from the National Natural Science Foundation of China (Grant Nos. 52071197, 52171174, and 52472274). H.Y. Chen acknowledges the support from the National Natural Science Foundation of China (Grant No. 52275467). C.D. Cao acknowledges the support from the National Natural Science Foundation of China (Grant No. 52271037) and Shaanxi Provincial Natural Science Fundamental Research Program, China (Grant No. 2023-JC-ZD-23).

\section*{Declarations}

\textbf{Conﬂict of interest}

The authors declare that they have no conﬂict of interest.

%
Kaitong Sun: Conceptualization, data curation, formal analysis, investigation, methodology, visualization, writing-original draft. 
Si Wu: Conceptualization, funding acquisition, methodology, project administration, supervision, visualization, writing-review \& editing. 
Guanping Xu: Data curation, formal analysis, investigation, methodology, visualization. 
Lingwei Li: Data curation, formal analysis, investigation, methodology, visualization. 
Hongyu Chen: Data curation, formal analysis, investigation, methodology, visualization. 
Qian Zhao: Data curation, formal analysis, investigation, methodology, visualization. 
Muqing Su: Data curation, formal analysis, investigation, methodology, visualization. 
Wolfgang Schmidt: Data curation, formal analysis, investigation, methodology, visualization. 
Chongde Cao: Conceptualization, funding acquisition, methodology, project administration, supervision, visualization, writing-review \& editing. 
Hai-Feng Li: Conceptualization, funding acquisition, methodology, project administration, supervision, visualization, writing-review \& editing.

\clearpage
%

\clearpage

\clearpage

\clearpage

\begin{table} [!t]
\caption{Refined structural parameters of an intermetallic ErPd$_2$Si$_2$ single crystal, including lattice constants, unit-cell volume (\emph{V}), atomic positions, isotropic thermal parameter $B_{iso}$, and reliability factors. These parameters were extracted from XRPD data collected at temperatures of 15, 135, and 285 K. The Wyckoff sites for all atoms are listed. The numbers in parentheses denote the estimated standard deviations of the (next-to-) last significant digit.}
\label{cellp}
\setlength{\tabcolsep}{11.8mm}{}
\renewcommand{\arraystretch}{1.1}
\begin{tabular} {lccc}
\hline
\hline
\multicolumn{4}{c} {A pulverized ErPd$_2$Si$_2$ single crystal (tetragonal, space group: \emph{$I4/mmm$})}                 \\
\hline
\emph{T} (K)                            &  15                              & 135               & 285                       \\
\hline
$a (= b)$ ({\AA})                       &  4.0881(1)                       & 4.0906(1)         & 4.0991(1)                 \\
$c$ ({\AA})                             &  9.8974(3)                       & 9.8878(3)         & 9.8808(3)                 \\
$V$ ({\AA}$^3$)                         &  165.41(1)                       & 165.45(1)         & 166.02(1)                 \\
$\alpha, \beta, \gamma$ $(^\circ)$      &  90, 90, 90                      & 90, 90, 90        & 90, 90, 90                \\
\hline
Er:                                     &  2\emph{a}: (0, 0, 0)                                                            \\
$B_{iso}$(Er) ({\AA}$^2$)               &  0.80(4)                         & 0.10(4)           & 1.73(4)                   \\
Pd:                                     &  4\emph{d}: (0, 0.5, 0.25)                                                       \\
$B_{iso}$(Pd) ({\AA}$^2$)               &  1.54(4)                         & 0.66(4)           & 2.51(4)                   \\
Si:                                     &  4\emph{e}: (0, 0, \emph{z})                                                     \\
\emph{z}(Si)                            &  0.3808(3)                       & 0.3757(4)         & 0.3794(4)                 \\
$B_{iso}$(Si) ({\AA}$^2$)               &  1.10(10)                        & 1.00(10)          & 2.90(10)                  \\
\hline
$R_p$                                   &  5.19                            & 5.32              & 5.55                      \\
$R_{wp}$                                &  8.26                            & 8.45              & 8.90                      \\
$\chi^2$                                &  3.02                            & 2.98              & 3.07                      \\
\hline
\hline
\end{tabular}
\end{table}

\clearpage

\clearpage

\clearpage

\clearpage

\clearpage

\clearpage

\clearpage

\clearpage

%

\end{document}